\documentclass[amsmath,amssymb,amsbsy,reprint,prl,preprintnumbers,showpacs,superscriptaddress]{revtex4-2}
\usepackage{graphicx,color}
\usepackage{dcolumn}
\usepackage{bm}
\usepackage{braket}
\usepackage{mathtools}
\usepackage{ulem}
\usepackage[breaklinks,colorlinks=true,linkcolor=blue,urlcolor=blue,citecolor=blue]{hyperref}
\usepackage{times}
\usepackage{physics}
\usepackage{latexsym}
\usepackage{amsmath,amssymb}
\usepackage{mathtools}
\usepackage{multirow}

\usepackage{graphicx}
\usepackage{bm,color}
\usepackage{physics}
\usepackage{amsmath, amssymb}
\usepackage{bm,color}
\usepackage{multirow}
\usepackage{ulem}

\newcommand{\DW}{{\rm DW}}
\newcommand{\oP}{{\, \rm op}}
\newcommand{\pE}{{\, \rm pe}}
\newcommand{\tw}{{\, \rm tw}}
\newcommand{\be}{\begin{eqnarray}}
\newcommand{\ee}{\end{eqnarray}}

\newcommand{\cmt}[2]{[#1,#2]}

\newcommand{\mi}{{\rm i}}

\newcommand{\zt}{{\mathbb{Z}_2}}
\newcommand{\uz}{U_{\mathbb{Z}_2}}
\newcommand{\ui}{U_{\mathbb{I}}}

\newcommand{\pX}[1]{X(#1)}
\newcommand{\pY}[1]{Y(#1)}
\newcommand{\pZ}[1]{Z(#1)}
\renewcommand{\pX}[1]{X_{#1}}
\renewcommand{\pY}[1]{Y_{#1}}
\renewcommand{\pZ}[1]{Z_{#1}}

\newcommand{\cf}[1]{c_{#1}}
\newcommand{\hdw}{H_{\rm DW}}
\newcommand{\mchsss}[6]
{
\left\{
\begin{array}{cc}
  #1 & #2   \\
#3 & #4   \\ 
#5 & #6
\end{array}
\right.
}

\begin{document}

\title{Topological domain-wall pump with $\mathbb{Z}_2$ spontaneous symmetry breaking} 
\date{\today}
\author{Yoshihito Kuno}
\thanks{The two authors contribute to the work equally.}
\affiliation{Graduate School of Engineering Science, Akita University, Akita 010-8502, Japan}
\author{Yasuhiro Hatsugai}
\thanks{The two authors contribute to the work equally.}
\affiliation{Department of Physics, University of Tsukuba, Tsukuba, Ibaraki 305-8571, Japan}

\begin{abstract}
A domain-wall pump
by an extended cluster model of $S=1/2$ spins is proposed
with local $U(1)$ gauge invariance.
Its snapshot ground state is gapped and doubly degenerated due to $\mathbb{Z}_2$ invariance,
which is broken by an infinitesimal boundary magnetic field.
  The ground state associated with the spontaneous symmetry breaking (SSB) is still symmetry-protected with additional spatial inversion that is characterized by the $\mathbb{Z}_2$ Berry phase.
We investigate the topological domain-wall pump with/without boundaries. The topological pump
associated with the inversion symmetry-breaking path
induces a non-trivial Chern number of bulk
and a singular behavior of edge states of the domain-wall.
Generalization to the multi-spin interaction is also explicitly given.
\end{abstract}


\maketitle

Experimental realization and simulation of Thouless pump \cite{Lohse2015,Nakajima2016,Lohse2018,Nakajima2021,Yatsugi2022,Stegmaier2024} sparked theoretical study
of topological pump that was
proposed three decades ago for non-interacting systems \cite{Thouless1983}.
It is natural to expect topological analogy for a system with many-body interaction \cite{TKNN1982,Niu1985}.
Effects of strong correlation for the
pump have been discussed intensively in recent theoretical studies
\cite{Berg2011,Rossini2013,Ke2017,Hayward2018,Nakagawa2018,Lin2020,Greschner2020,KH2021_1,Aligia2023}.
Also experimental study was recently reported \cite{Viebahn2024}. Non-linear effects of the topological pump have been also observed in a photonic experimental system \cite{Jurgensen2021}.

As for the correlated system, various degrees of freedom are coupled with each other and
some generic extended scheme beyond Thouless' first proposed is needed.
Although various theoretical studies on this issue have been reported, there is still no end to predict the existence of various types of topological pumps.
Recently, the authors proposed a general construction scheme \cite{Hatsugai2023}.
The key ingredient is a local $U(1)$ gauge symmetry and
a gapless critical point between
the symmetry-protected topological phases (SPTs)
that is a topological obstruction to guarantee
the pump is topologically non-trivial.
Using the critical point, inclusion of symmetry-breaking perturbations as
a synthetic dimension implies a non-trivial pump, where the pumping path wraps the symmetry protected gapless point without gap closing \cite{Hatsugai2023}.
This guiding principle is general and the scheme opens the door for the construction of a variety of topological pumps.

In this letter, we propose a domain-wall pump where
the gauge symmetry is associated with the domain-wall as a basic degree of freedom
of $S=1/2$ quantum spins.
It is an extended cluster model with modulated on-site interaction. 
The ground state of the Hamiltonian is doubly degenerated due to $\mathbb{Z}_2$ symmetry,
which is broken spontaneously.
This is a non-trivial topological pump of the SSB state.

We focus on the open boundary condition \cite{Verresen2022}
and consider an extended cluster spin model of $S=1/2$ quantum spins
defined on $(L+1)$-sites (we set $L\in 4\mathbb{Z}$.) \cite{Suzuki1971,Raussendorf2003,Son2011,Smacchia2011,Nie2017}
\begin{eqnarray*}
  \hdw^{\oP} &=&-\sum^{L-1}_{j=1}J_j(\pX{j}-\pZ{j-1}\pX{j}\pZ{j+1})
  + \sum_{j=0}^{L-1}\Delta _{j+\frac 1 2 }\pZ{j}\pZ{j+1},
\label{h0}
\end{eqnarray*}
where $\pX{j},\pY{j},\pZ{j}$ are the Pauli matrices ($\pX{j}\pY{j}=\mi \pZ{j}$, etc.).
The first term of $\hdw^{\oP}$ represents a domain-wall hopping and the second term is the Ising interaction regarded as an on-site potential of the domain-wall. The Hamiltonian $\hdw^{\oP}$ is  ${\mathbb{Z}_2}$ invariant by a parity 
$U_{\zt}Z_jU^\dagger_{\zt}=-Z_j$, $U_{\zt}=\prod_{j=0}^L\pX{j}$, $\cmt{\hdw^{\oP}}{U_{\zt}}=0$.

The Hamiltonian $\hdw^{\oP}$ is an extension of the cluster model or the transverse field Ising model.
The physical picture is clear by writing it with a set of new Pauli matrices
on the link due to the KW transformation \cite{Savit1980,Kohmoto1981,Linhao_Li2023},
$\pZ{j} = {\prod}_{k=j}^{L}\pX{k+\frac 1 2 }$, ($j=0,\cdots,L$),
$\pX{j} =
{\pZ{\frac 1 2 }}{(j=0)},
{\pZ{j-\frac 1 2 }\pZ{j+\frac 1 2 }}{(j=1,\cdots,L)}$.
Its inverse is given by
$Z_j=\prod^{L}_{k=j}X_{k+\frac{1}{2}},\:\:(j=0,\cdots ,L)$,
$X_0=Z_{\frac{1}{2}}$, $X_j=Z_{j-\frac{1}{2}}Z_{j+\frac{1}{2}}$ ($j=1,\cdots,L$).
Then, by using the subsequent Jordan-Wigner transformation to the canonical fermion $c_j$,
$
\pZ{j+\frac 12 } =
-\mi e^{-i\pi\sum_{k=1}^{j-1} n_{k+\frac 1 2 }}
(c_{j+\frac 1 2} ^\dagger- c_{j+\frac 1 2})$,
$\pX{j+\frac 1 2 } =  1-2 n_{j+\frac 1 2}$ and
$n_{j+\frac 1 2}=\cf{j+\frac 1 2}^\dagger \cf{j+\frac 1 2}$. The Hamiltonian $\hdw^{op}$ is mapped into a YZ model with an extra single free spin
(the Rice-Mele model with an extra free fermion). 

\begin{figure}[t]
\begin{center}
\vspace{0.5cm}
\includegraphics[width=7.5cm]{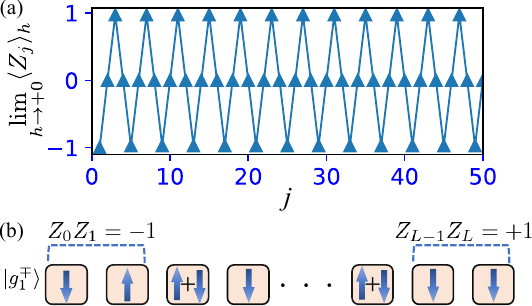}
\end{center}
\caption{
  (a) 
  Local magnetization $\lim_{h\to +0} \langle \pZ{j}\rangle _h$ for $J=1$, $\delta J=0.5$ (away from the strong coupling limit) is shown for 
   $1\le j\le 50$ for the system with $512$ lattice sites under open boundary condition.
  (b) One of the non-trivial SPT state with domain-wall edge states at the left boundary
  $|g^{\mp}_1\rangle$. The up and down blue arrows represent the z-directional spin.
}
\label{Fig0}
\end{figure}

Note that the Hamiltonian $\hdw^{\oP}$ has double degenerate groundstates and exhibits a ${\mathbb{Z}_2}$ spontaneous symmetry breaking (SSB).
These doubly degenerate groundstates are labeled by a boundary magnetization, that is, the boundary spins 
$\pZ{0}$
and $\pZ{L}$
commute
with the Hamiltonian $\hdw^{\oP}$, $[\hdw^{op},Z_{0(L)}]=0$. 
By considering the Hamiltonian $\hdw^{\oP}+hZ_{L}$ (where $h$ is a local magnetic field), we have shown the magnetization is finite, $\lim_{h\to +0}\langle Z_j\rangle_{h}\neq 0$ (SSB), on a case which we mainly focused. 
Assuming the coupling $J_j$ is positive and modulated as, $J_j= J-(-1)^j\delta J$, $J>0$
and $|\Delta _{j+\frac 1 2}|\ll |\delta J|$, the ground state is a period-4-antiferromagnetic and gapped
except for $\delta J=0$. For $\delta J\lesssim J$, the bulk exhibits the SSB as
$
|g\rangle \simeq \cdots|\downarrow\rangle_{j-1}|+\rangle_{j}|\uparrow\rangle_{j+1}\cdots
$
where $\pZ{j} |\uparrow\rangle_j =| {\uparrow} \rangle_j$,
$\pZ{j}|\downarrow\rangle_j =-| {\downarrow} \rangle_j$
and $X_j|+\rangle_j=|+\rangle_j$.
In fact, we have numerically confirmed this non-zero spontaneous magnetization
as shown in Fig.\ref{Fig0}(a),
$\lim_{h\to +0} \langle \pZ{j}\rangle _h=
\lim_{h\to+ 0} \langle \pX{j+\frac 1 2 }\cdots\pX{L+\frac 1 2 }\rangle _h =- \det \psi_G^\dagger D_j\psi_G$,
where the fermionic ground state is $|G \rangle =\prod_{\ell=1}^M (\bm{c}^\dagger \psi_\ell )|0 \rangle=|\psi_G \rangle$, $\psi_G\equiv(\psi_1,\cdots,\psi_M)\in M(L,M) $ ($M$ is a number of negative energy fermionic states) and
$D_{j}={\rm diag }\, (1,\cdots,1,-1,\cdots,-1)$ is an $L$ dimensional diagonal matrix (the first $j$-th elements are $+1$)
due to the Jordan-Wigner transformation mentioned above.

With the open boundary condition, 
the domain-wall current 
averaged over the system is given as
$ 
  {\cal J} =  \hbar ^{-1} \partial_{\theta}\hdw^{\oP,\theta }
$ 
where $\hdw^{\oP,\theta}=U_L \hdw^{\rm op}U_L ^\dagger $ 
and
the large gauge transformation $U_L=e^{-\mi \theta {\mathcal P}}$ is generated by the center of mass (CoM)
of the domain-wall
$\mathcal{P}=-\sum_{j=1}^{L-1}x_{j+\frac 1 2 }\pZ{j}\pZ{j+1}/2$ \cite{Hatsugai2016,localG}.
The normalized position of the $j$-th link measured from
the center, $x_0=L/2$, is $x_{j+\frac 1 2 } =(j+\frac 1 2-x_0)/L \in (-1/2,1/2)$.
Note that
$\hdw^{\oP,\theta}$ is obtained from
$\hdw^{\oP}$ by replacing all $J_j\to \hat J_j\equiv J_j e^{-\mi \frac {\theta }{L} \pZ{j}\pZ{j+1}/2}$
(This is an operator valued phase factor).
It implies $j\equiv \langle G|\mathcal{J}|G\rangle =\partial_t  \langle G| \mathcal{P} |G \rangle $
where $|G \rangle $ is a many-body wave function associated with the 
the Schr\"odinger equation,
$\mi \hbar  \, \partial_t |G \rangle =\hdw^{\oP}|G \rangle $.
This is generically valid for the time-dependent Hamiltonian, that is,
the time dependent (a pump as periodic in time)
$J_j$ and $\Delta _j$ and independent of the boundary condition.
Also the Hamiltonian $\hdw^{\oP}$ is $U(1)$ invariant, $\cmt{\hdw^{\oP}}{N_D}=0$
where the domain-wall charge is given by
\begin{align*}
  N_D &= -\sum_{j=0}^{L-1}\pZ{j}\pZ{j+1}/2. 
\end{align*}
 It prevents from pair annihilation of the domain-wall. We restrict to the $N_D=0$ sector in the following discussion.

Assuming the system is gapped and the time dependence of the parameters are slow enough,
the  current is evaluated by the adiabatic approximation as
$j\simeq j_A^{\oP}=\partial_t P^\oP$, $P^\oP =\langle \mathcal{P} \rangle^{\oP} $
where $\langle \cdot \rangle^{\oP} $ is a average over the degenerated ground states
of the snapshot Hamiltonian $\hdw^{\oP}$ (trace over the degenerate multiplet divided by the degeneracy).
When the system is topologically non-trivial, we may expect an existence of the edge states, which implies
the system is gapless and breakdown of the adiabatic approximation. Due to the singularities, $P^\oP$ is generically discontinuous at $t=t_1,t_2,\cdots$,
associated with the jump $\Delta P(t_k)$.
Then the pumped domain-wall charge during the period $T$, $Q^\oP=\int_0^Tdt\, j_A^\oP$,
is given by the minus of the total jumps, $I_\DW=-\sum_k \Delta P^\oP(t_k)$
as $Q^\oP=I_\DW $ since the adiabatic ground state returns to the same one after the cycle that implies
compensation of the pumped charge by the discontinuities.
This is quantized by integers due to the $U(1)$ invariance in the large system size limit, $L\to\infty$ \cite{Hatsugai2016,Hatsugai2023,Kuno2021_v2}.

Since the quantization of the discontinuities $I_\DW$ 
implies the edge states are stable for the
continuous deformation of the Hamiltonian, let us discuss the domain-wall edge states
in the strong coupling limit. 
In contrast to the well known cases such as the dimerized chains
(quantum disordered chains),
the strong coupling limits of the present case 
possess explicit SSB.
By setting $\Delta _{j+1/2}=(-1)^j\Delta$ and $|\Delta| \ll |\delta J|$.
Here we consider two strong coupling limits with/without domain-wall edge states.
For $J_{j \in \rm even}=0$ ($\delta J =J$), we may take the doubly degenerate ground states as
\begin{eqnarray}
|g_1\rangle &=& |\uparrow\rangle_{0}|+\rangle_{1}|\downarrow\rangle_{2}|+\rangle_{3}|\uparrow\rangle_{4}\cdots |\downarrow\rangle_{L-2}|+\rangle_{L-1}|\uparrow\rangle_{L} ,\nonumber\\
|g_2\rangle &=&
|\downarrow\rangle_{0}|+\rangle_{1}|\uparrow\rangle_{2}|+\rangle_{3}|\downarrow\rangle_{4}\cdots  |\uparrow\rangle_{L-2}|+\rangle_{L-1}|\downarrow\rangle_{L}.\nonumber
\end{eqnarray}
They are associated with SSB and apparently domain-wall charge is vanishing $\langle \pZ{j}\pZ{j+1} \rangle^\oP =0$, $j=0,\cdots,L-1$.
The phase is topologically trivial.
As for $J_{j \in \rm odd}=0$ ($\delta J =-J$), the ground state multiplet is spanned by the two states with domain-wall at the both boundaries
\begin{eqnarray}
  |g^{\mp}_1\rangle &=& |\downarrow\rangle_{0}|\uparrow\rangle_{1}|+\rangle_{2}|\downarrow\rangle_{3}|+\rangle_{4}\cdots |+\rangle_{L-2}|\downarrow\rangle_{L-1}|\downarrow\rangle_{L}, \nonumber\\
  |g^{\mp}_2\rangle &=&
|\uparrow\rangle_{0}|\downarrow\rangle_{1}|+\rangle_{2}|\uparrow\rangle_{3}|+\rangle_{4}\cdots |+\rangle_{L-2}|\uparrow\rangle_{L-1}|\uparrow\rangle_{L}.\nonumber
\end{eqnarray}
for $\Delta >0$ (See Fig.~\ref{Fig0}(b))
and
\begin{eqnarray}
  |g^{\pm}_1\rangle &=& |\uparrow\rangle_{0}|\uparrow\rangle_{1}|+\rangle_{2}|\downarrow\rangle_{3}|+\rangle_{4}\cdots |+\rangle_{L-2}|\downarrow\rangle_{L-1}|\uparrow\rangle_{L}, \nonumber\\
  |g^{\pm}_2\rangle &=&
|\downarrow\rangle_{0}|\downarrow\rangle_{1}|+\rangle_{2}|\uparrow\rangle_{3}|+\rangle_{4}\cdots |+\rangle_{L-2}|\uparrow\rangle_{L-1}|\downarrow\rangle_{L}.\nonumber
\end{eqnarray}
for $\Delta <0$.
Although the states are associated with SSB,
the local domain-wall charge is independent of the SSB pattern and  is given as
$\langle \pZ{j}\pZ{j+1} \rangle^\oP =
-{\rm sgn}\, \Delta\, (j=0)$,
${\rm sgn}\, \Delta\, (j=L-1)$, and  $0$ ($j$: otherwise).
It implies the phase is topologically non-trivial.
During the adiabatic time evolution, the sign change of the $\Delta$ from the positive to the negative
induces the discontinuity of the CoM, $\Delta P^\oP=-1$, and contributes to $I_\DW$ by $1$. 
The two phases $J_{\rm even}\ll J_{\rm odd}$
and
$J_{\rm even}\gg J_{\rm odd}$ when $\Delta =0$ are the SPT phases
protected by the site-center inversion, 
  $\ui$, $\ui \sigma_j \ui^\dagger= \sigma _{L-j}$, $\sigma_j=\pX{j},\pY{j},\pZ{j}$,  $j=0,\cdots,L$ \cite{sinv} 
where the topological order parameter is the $\mathbb{Z}_2$ Berry phase defined by the site twist (See below).

As for the periodic boundary condition, the Hamiltonian $\hdw^{\pE,\theta }$,
which is translationally invariant, is
obtained from $\hdw^{\oP,\theta}$ by identifying the sites $j=0$ and $j=L$.
Then the twisted Hamiltonian at the 
site $L\equiv 0$ is defined by following the above definition
as $\hdw^{\pE,\theta }=U_L \hdw^{\tw,\theta} U_L ^\dagger $.
The $\theta $ dependent twist term of $\hdw^\tw$ is explicitly written as
\begin{align}
h_ \theta =-J_0 \big((\pX{0}-\pZ{L-1}\pX{0}\pZ{1})\cos \theta
+(\pZ{L-1}\pY{0}-\pY{0}\pZ{1}) \sin \theta \big).\nonumber
\end{align}
This twist term is also invariant for the site-center inversion 
$\ui:j\to -j$ (mod\, $L$) supplemented by $\theta \to - \theta $
and also $\mathbb{Z}_2$ invariant, 
$\cmt{
  \hdw^{\tw,\theta}
  }{U_{\mathbb{Z}_2}}=0$.
Noting that $\uz$ is invariant for the site inversion,
we can define $\mathbb{Z}_2$ Berry phases for the invariant subspaces of the eigenspace of $\uz$, as \cite{z2}, 
$\gamma^\pm=-\mi \int_0^{2\pi} d \theta \, A^\pm_ \theta =0,\pi\ (\text{mod }\, 2\pi)$ 
where 
$A^\pm_\theta   = 
\langle g^\pm| \partial_{\theta } g^\pm \rangle $, $\hdw^{\tw,\theta }|g^\pm \rangle =|g ^\pm \rangle E$,
$\uz | g^\pm \rangle =\pm | g^\pm \rangle $ and
$\langle g^\alpha | g^\beta \rangle =\delta^{\alpha \beta }$, 
($\alpha, \beta =\pm$)
\cite{Hatsugai2004,Hatsugai06}.
This $\mathbb{Z}_2$ quantization guarantees
existence of the two SPT phases for $\delta J>0$ and $\delta J<0$ and the critical points between them. 
In the strong coupling limit $\delta J=-J$, $J_{\rm odd}=0$, the system
is a collection of the decoupled trimers. 
Then the ground state is 
given by $| g _\pm \rangle =|\Psi_\pm\rangle \otimes|\psi_\pm \rangle $
where $h_ \theta |\Psi_\pm \rangle=-2J_0|\Psi_\pm \rangle $, and
 $|\psi_\pm \rangle $ is a $\theta $-independent state of the spins at $j=2,\cdots,L-2$. 
Explicitly it is written as 
$|\Psi_\pm \rangle=(|A_{\pm}\rangle e^{\mi \theta }+|B_{\pm}\rangle)/{2} $ where 
$|A_{\pm}\rangle=|\uparrow\downarrow\downarrow\rangle
\pm |\downarrow\uparrow\uparrow\rangle)
$,
$|B_{\pm}\rangle=|\uparrow\uparrow\downarrow\rangle
\pm|\downarrow\downarrow\uparrow\rangle 
$ and 
$\gamma _ \pm =\pi$ \cite{Hatsugai2004,Hatsugai06}. This non-trivial $\mathbb{Z}_2$ quantization is 
adiabatically stable for inclusion of interaction between the trimers. 
Since the domain-wall edge states is also stable as far as the bulk is gapped, 
this $\mathbb{Z}_2$ quantized Berry phase $\gamma_\pm$ is
a topological order parameter.
Non-trivial $\gamma _\pm =\pi$ implies that the domain-wall edge states appear
at the boundaries where the twisted trimer is broken.

  
The existence of the two SPT phases at $\Delta_{j+/1/2} =0$ also implies an existence of non-trivial topological pump passing around the
gap closing point between the two SPTs.
By introducing the symmetry breaking term $\Delta\ne 0$, we construct a topological pump
of the domain-wall.
Explicitly, we set the time dependence of the parameters
as
$J_j=[1-(-1)^j \delta J \sin (2\pi t/T)]$ and
$\Delta_{j+1/2}=(-1)^j \Delta _0 \cos (2\pi t/T)$ where $T$ is
a period.

In contrast to the open boundary condition, the current under the adiabatic approximation
$j_A^\pE$ is weakly $\theta $ dependent.
Since its dependence is small as ${\cal O} (e^{-L/\xi})$, $\xi>0$ for a large system\cite{Kudo19}, 
the twist averaged current $\bar j_A^\pE=\int_0^{2\pi}\frac{d \theta}{2\pi} \, {j_A^\pE}$
is useful.
It gives the pumped charge of the periodic system $Q^\pE=\int_0^Tdt\, \bar j_A^\tw=C_\DW$
where $C_\DW=\frac 1 {2\pi \mi} \int_0^Tdt\,\int_0^{2\pi}d \theta \, B$ is 
the Chern number associated with the normalized Berry connection of the degenerate ground state multiplet.
To be explicit, it is given as
$B=\partial_\theta  A_t-\partial_t A_ \theta $,
$A_  \mu =(1/2) {\rm Tr}\, \psi ^\dagger \partial_ \mu \psi$
\cite{Kudoadia21}
where
$\psi=(|g _1 \rangle ,| g_2 \rangle )$ is the degenerate groundstate
doublet where $| g _i \rangle $'s are any orthonormalized degenerate ground states
( $\langle g_i|g_j \rangle =\delta _{ij}$).
Taking the orthonormalized multiplet as 
$\psi=(|g_S \rangle , |g_A \rangle )$ where 
$|g_S \rangle =(|g^+ \rangle + |g^- \rangle )/\sqrt{2} $ and $|g_A \rangle =\uz | g_S \rangle= (|g^+ \rangle - |g^- \rangle )/\sqrt{2} $, the Berry connection is given as $A_ \mu = \langle g_S| \partial_\mu g_S \rangle $ since $\uz$ is parameter independent. It implies $C_{\rm DW}\in \mathbb{Z}$.
Since we may expect $j_A^\oP=j_A^\pE$ for the infinite system, the bulk-edge correspondence
of the domain-wall pump is written as
\begin{align*}
  I_\DW&= C_\DW.
\end{align*}

Let us discuss a generalization of the DW pump
by introducing the $k$-string \cite{Suzuki1971,Tantivasadakarn2022} that is
a bilinear majorana operator via the Jordan-Wigner transformation as
\begin{align*}
  h_j^k &= \mchsss
  {\pY{j+k}(\prod_{\ell=j+k+1}^{j-1}\pX{\ell})\pY{j}\equiv
  {\text {Str}}^Y_{j+k,j}
  }{k<0}
  {-X_j
  }{k=0}
  {\pZ{j}(\prod_{\ell=j+1}^{j-1+k}\pX{\ell})\pZ{j+k}\equiv
  {\text {Str}}^Z_{j,j+k}
  }{k>0},
\end{align*}
where $[h^{k}_i,h^{k}_j]=0$ for $\forall i,j$ and $k \in \mathbb{Z}$.
It gives
$
  (h_j^k)^2 = 1
$,
$  \pX{j}h_j^k\pX{j+k} = h_{j+k}^{-k}$ and the pivoting (Unitary transformation)\cite{Tantivasadakarn2022}
\begin{align*}
  U_k h_j^{k_0} U_k ^\dagger &= h_{j+k_0-k}^{2k-k_0}
\end{align*}
where $U_k=\prod_{j=-\infty}^\infty e^{\mi\pi/4}(1-\mi h_j^k)/\sqrt{2}$ and $U_k^2=1$. 
Noting that the domain-wall hamiltonian of the open boundary condition is written as
$\hdw^\oP(0,1,\cdots,L)$$ =\sum_{j=1}^{L-1}J_j(h_j^0+h_{j-1}^2)+\sum_{j=1}^{L-1}\Delta _{j+1/2} h_j^1$
and $N_D(0,1,\cdots,L)=-\sum_{j=0}^{L-1}h_j^1/2$ where
relevant sites are explicitly specified,
the length $2\ell$ extended domain-wall hamiltonian is defined by 
\begin{align*}
& \hdw^{\oP,\ell} (0,1,\cdots,L) \equiv U_{\ell} \hdw^{\oP}( \ell-1,\cdots, L-\ell +1 ) U_{\ell} ^\dagger
\\
&= 
    \sum_{j=\ell}^{L-\ell} 
      J_j({\rm Str}_{j-\ell,j+\ell}^Z
+ {\rm Str}_{j+1-\ell,j-1+\ell}^Z)
\\
&+ \sum_{j=\ell-1}^{L-\ell}\Delta_{j+1/2}
{\rm Str}_{j+1-\ell,j+\ell}^Z.
\end{align*}
It implies that $2(\ell-1)$ spins at the following sites are free  before the pivoting,
$ (0,1,\cdots,\ell-2)$ and $(L-\ell+2,\cdots,L) $.
It implies the degeneracy
$2\cdot 2^{2(\ell-1)}=2^{2\ell-1}$ for all states.
It commutes with $N_D^\ell =-\sum_{j=\ell-1}^{L-\ell} {\rm Str}_{j+1-\ell,j+\ell}^Z/2$.
We may also consider a pivotting after the Kramers-Wannier transformation.
Another possible extension is interacting domain-wall pump. 
One of the possible interactions among
the $M$ domain-walls that is compatible with the $U(1)$ invariance can be defined as
\begin{align*}
  H_{\rm int} &= V \sum_{j}  h_{j}^1\cdots h_{j+M}^{1}.
\end{align*}
As far as the interaction is small compared with the gap, the topological domain-wall pump
that is governed by $\hdw+H_{\rm int}$ is stable. 

As a summary, we proposed a topological domain-wall pump with SSB. It is 
an extended cluster model of the quantum spins where creation and annihilation of the
domain-wall are prohibited. It results in the $U(1)$ invariance
generated by the domain-wall charge and the conservation of domain-walls.
The center of mass of the domain-wall weighted by the link position is
the generator of the large gauge transformation that induces a topological current of the domain-wall.
The gauge field lives on the sites in contrast to the standard topological pump of the charge.
This pump is novel and non-trivial, that is, the SSB and topological pump coexist. The bulk-edge correspondence of the domain-wall pump is discussed by using edge states of the domain-walls and the Chern number of the bulk.
Generalization of the domain-wall pump is also proposed by the pivoting and inclusion of the domain-wall interaction.

{\it Acknowledgements.---}
The work is supported by JSPS
KAKEN-HI Grant Number JP21K13849 (Y.K.),
23K13026 (Y.K.), 23H01091(Y.H.), 23K25788 (Y.H.) and
JST CREST, Grant No. JPMJCR19T1 (Y.H.), Japan.\\

\bibliography{ref-yh}

\begin{thebibliography}{40}
\bibitem [{loc()}]{localG}%
  \BibitemOpen
  \href@noop {} {\ }\bibinfo {note} {$ U \hdw^{\oP}(\{ J_{j}\}) U ^\dagger =\hdw^{\oP}(\{ \hat J_{j}\}) $ with $\hat J_j= J_j e^{\mi \Theta}$, $\Theta=\mi \pZ{j}\pZ{j+1}(\theta _{j- 1 /2 }-\theta _{j+ 1/ 2 })$, for the gauge transformation, $ U = e^{\mi \Omega}$, $\Omega=\sum_{j=0}^{L-1} \theta _{j+ 1/ 2 } Z_j Z_{j+1}/2 $. In contrast to the standard gauge transformation, $\Theta $ is not a simple phase but an operator valued.}\BibitemShut {Stop}%
\bibitem [{sin()}]{sinv}%
  \BibitemOpen
  \href@noop {} {\ }\bibinfo {note} {$U_{\mathbb{I}}= \prod_{j=0}^{L/2-1}P_{j,L-j}$ where $P_{ab}=(1+ \pX{a}\pX{b}+ \pY{a}\pY{b}+ \pZ{a}\pZ{b})/2$ is the exchange of the spins at $a$ and $b$, $P_{ab}\pX{a}P_{ab}=\pX{b}$, $P_{ab}\pY{a}P_{ab}=\pY{b}$, and $P_{ab}\pZ{a}P_{ab}=\pZ{b}$ for the open boundary condition and $U_{\mathbb{I}}= \prod_{j=1}^{L/2-1}P_{j,L-j}$ for the periodic/twisted boundary condition.}\BibitemShut {Stop}%
\bibitem [{z2()}]{z2}%
  \BibitemOpen
  \href@noop {} {\ }\bibinfo {note} {Noting that $ \cmt{\ui}{\uz}=0$, $\uz (\ui| g^{\pm} \rangle )= \pm(\ui | g^{\pm} \rangle )$ and $\ui |g ^{\pm} \rangle =|g^{\pm} \rangle e^{\mi \Omega_\pm } $. Also $\ui \hdw^{\tw,\theta } \ui = \hdw^{\tw,-\theta }$, $\ui| g^{\pm} (\theta )\rangle \propto | g^{\pm}(-\theta ) \rangle $. It implies $A_ \theta ^\pm=-A _ \theta ^\pm+ \mi \partial_ \theta \Omega$ and $\gamma ^\pm \equiv-\gamma ^\pm,\ \text{mod}\, 2\pi$.}\BibitemShut {Stop}%
\end{thebibliography}

\end{document}